# Success in IT offshoring: Does it depend on the location or the company?

*Jongkil Jeong*

*Abstract* – **Many companies are now looking towards offshore vendors to fulfill their outsourcing requirements. With this growth, we have seen particular countries such as India dominate the offshoring market, and this paper will examine what type of role the reputation of a particular offshoring country has on the decision making process of firms looking to offshore.**

## I. INTRODUCTION

Offshoring is defined as the relocation of a particular business process to lower-cost locations outside national borders. (Erber & Sayed-Ahmed, 2005)

Outsourcing services to offshore providers has become increasingly popular over the past few decades, with the offshore information services market alone estimated to be valued at $30 billion to $40 billion globally. (Delmonte A. & McCarthy R, 2003)

The reasons why companies choose to offshore have been well researched and documented. Companies offshore as they are looking to lower costs, increase productivity, increase the quality of work and access skilled labour markets (Kliem 2004).

However, the most important factor out of these benefits is cost savings. These cost savings are primarily associated with lower labour costs for highly qualified staff that are readily available in different countries (e.g. Mahnke 2006; Lewins & Peeters 2006).

Despite the significant cost savings firms can receive from lower labour costs, many companies still face various challenges and risks. Furthermore, additional challenges unique to offshoring have also been identified in addition to the risks experienced in outsourcing in general (Kliem 2004).

These additional risks and challenges have resulted in an estimated 50% failure rate of offshored IT projects (McCue 2005) and as a result, companies are now "backsourcing" their IT services – a terminology referring to terminating or not renewing an outsourcing agreement so that functions can be brought back in-house (Cutter Consortium 2005). Also, there has been research to suggest that a high proportion of companies are not ready to try offshore development at all. (Bowers, 2008)

To mitigate some of these risks and challenges, both corporate entities and academics have conducted various researches to determine ways to choose the appropriate offshore vendor to ensure the chance of success when offshoring is increased.

The conclusions and findings from the various research papers have resulted in different models and theories to be designed and developed. These theoretical models and frameworks can generally be categorized into two different opinions as to how to best overcome the barriers of successfully outsourcing, and to increase the likelihood of success by choosing the appropriate offshore vendor.

On one hand, there have been studies that have suggested that issues faced with offshoring can be best resolved by offshoring to companies in specific countries which have become specialists in a particular service area (e.g. Dossani & Kenney 2007).

For example, countries such India, Russia, Eastern Europe and China have become well-known destinations to offshore IT and knowledge based services, whereas the Philippines has become a global hotspot for firms looking to offshore their call center operations due to the high percentage of skilled works being able to speak both fluent English and Spanish which provides the country a distinct advantage when servicing consumers. (Manning et al. 2009)

These papers state that by choosing the appropriate offshore vendor based on geographical location, a company is able to drastically improve the chances of success when outsourcing to an offshore provider.

On the other hand, certain parties have focused on trying to find a solution by approaching the issue from a business perspective. These parties state that geographical & cultural elements do not pose a major risk, and that success in offshoring depends on having the right planning, management and execution to eliminate financial, technical, managerial, behavioral and legal risks (e.g. Kliem 2004).

Rather than choosing an offshoring vendor based on geographical location, these parties conclude that the success of offshoring highly depends on the capacity, experience and knowledge that the offshore vendor has in regards to the type of service the company looking to offshore requires.

Their main argument is that IT services & development is becoming more of a commodity rather than a service, similar to manufacturing industries. Harrison & McMillan (2006) argues that IT is likely to become similar to manufacturing industries such as automotive and textile, where production lines will be formed to provide simple coding of well specified programs that can be developed similar to commoditized tasks in manufacturing industries (Carr, 2003).

In this regards, these scholars believe that choosing an IT vendor with the experience and capacity to deliver on their services are the most vital elements when creating a successful offshore model as well as proper planning and management by the firm wishing to offshore.

In this paper, we will critically examine evidence which support both sides of the argument. We will identify some of the key elements which have been presented to support either sides of the argument, and summarize the main points drawn to come to a conclusion as to whether or not the geographical location of a company has an effect on increasing the likelihood that an offshore project will result in success for a particular firm.

## II. DOES LOCATION MATTER?

It is a well-known fact that certain geographical locations are more preferred than others when offshoring IT services.

Certain parties agree that determining the offshoring location is deemed to be a key aspect of making decisions regarding offshoring, and the rationale for selecting particular locations over others has been researched by a number of studies in recent years (e.g. Manning et al. 2009).

Recent studies suggest that certain geographical areas have become experts in a specific service field over time. For example, India has become a well renowned destination for firms wishing to outsource their IT or Back-office services, whereas the Philippines has become renowned for providing call center services (Lewin & Peeters 2006).

As seen in Table 1, India is by far the most preferred location for business process and IT applications, followed by China, and other nations such as the Philippines, Mexico, Latin America and Canada. It is also interesting to note that "Other Asian Countries" (Indonesia, Malaysia, Vietnam etc.) make up 7% of the overall locations in offshoring as well.

| Locations | % of existing implementations | % of new implementations (next 18 to 36 months) |
|---|---|---|
| India | 69% | 66% |
| China | 7% | 7% |
| Other Asia | 7% | 16% |
| Latin America | 6% | 1% |
| Philippines | 4% | 3% |
| Canada / Mexico | 4% | 1% |
| Eastern Europe | 3% | 6% |

Table 1: Location of Offshoring (Lewin & Peeters 2006)

So why are firms looking at offshoring their services to specific countries like India, rather than other countries around the world?

To answer this question, we must understand some of the factors firms consider when looking for an offshore vendor.

Lewins & Peeters (2006) conducted research as to see what the primary drivers for firms to offshore were and came up with the following results:

| Strategic Drivers | % of respondents citing driver as important |
|---|---|
| Taking out cost | 93% |
| Competitive pressure | 69% |
| Improving service levels | 56% |
| Accessing qualified personnel | 55% |

Table 2: Strategic Drivers of Offshoring (Lewins & Peeters 2006)

Based on Table 2, we are able to see that there needs to be a distinct advantage in labour costs due to competitive pressure in order for firms to offshore. The primary driver firms look to offshore their work is to reduce the high labour costs associated with hiring IT specialists, so they may stay competitive in the market.

Table 3 shows the average salaries of a programmer in relative countries in 2003, and the projected increase it is expecting by 2015.

| Projected Programmer Annual Salaries | | | |
|---|---|---|---|
| | 2003 | 2015 | % Change |
| US | $74,486 | $85,000 | 14% |
| Japan | $30,338 | $35,000 | 15% |
| Malaysia | $6,930 | $9,000 | 30% |
| Philippines | $6,549 | $9,000 | 37% |
| Germany | $39,879 | $65,000 | 63% |
| China | $5,852 | $10,000 | 71% |
| England | $38,450 | $67,000 | 74% |
| France | $37,250 | $65,000 | 74% |
| India | $6,350 | $20,000 | 215% |
| **Russia** | $7,540 | $25,000 | 232% |
| **Thailand** | $1,760 | $8,000 | 355% |
| **Poland** | $8,990 | $45,000 | 401% |

**Table 3: Projected Programmer Annual Salaries (Ferguson 2004)**

As you can see from the average salaries in the relevant countries, you can see that there is a strong correlation between the average salaries in 2003 and the actual destination of offshoring as seen in Table 1.

This proves that location does play a critical part when firms are looking at offshoring, as the country they wish to offshore to, has a direct correlation with the cost reductions they anticipate from saving on labour costs.

The second main driver firms look to offshore is to enhance productivity & improving service quality.

In the paper written by Gupta et al (2006), they describe the concept of the 24-hour knowledge factory, and explained that firms are able to benefit from strategic offshoring as it allows firms to offer services, and work continuously around the clock.

They state that offshoring allows firms to create a global network of teams, who are able to work on specific projects around the clock, with each team in a different geographical location working in their allocated time zones which will resulted in drastically reduced turnaround times for projects, as well as allow firms to provide around the clock service and support to their clients (Gupta et al 2006).

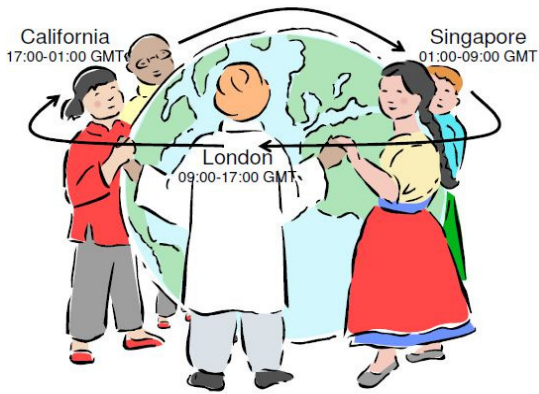

**Diagram 1: 24-hour global knowledge factory (Gupta et al 2006)**

This around the clock operation as demonstrated in Diagram 1, is only possible if firms choose carefully their offshore destinations, and further proves the point that location does matter when trying to offshore.

The third main driver as described in Table 1 is firms looking to access qualified, skilled personnel. The shortage of skilled professionals locally coupled with finding motivated and highly qualified staff willing to take on low-level jobs (Lewins & Peeters 2006) has resulted in firms looking at destinations where they are able to fulfill this requirement.

To provide an example, the US has been experiencing a declining number of students enrolling in Information Technology & Information Systems courses in tertiary institutions. This has created issues for firms as there are insufficient numbers of IT / IS graduates which are forcing companies to move to markets where this shortage of labour can be fulfilled (Ives 2005).

This has naturally opened up opportunities for countries with an abundance of skilled, educated workers such as India to provide firms the necessary manpower to provide the service it requires. Furthermore, Indian vendors have the potential of growing offshore centers to thousands of IT professionals if desired, due to the access the country has to a highly educated labour force (Carmel & Agarwal 2002).

In summary, these views show that there needs to be certain characteristics a country needs to have in order for it to be a feasible choice for firms wishing to offshore; (a) Labour Costs must be cheaper than the firms country of origin; (b) The offshore vendor must be in a geographical location which allows increased productivity and service continuity; and (c) The country must possess highly skilled & qualified workers.

Although there were other reasons as to why companies were looking to offshore, these three main drivers explain as to why the location of the offshoring vendor plays such a significant part of the business decision making process for firms.

### III. DOES THE COMPANY MATTER?

On the other hand, there are specific parties who do not consider location to be the most important factor when choosing an offshore IT vendor.

Some parties say that this is reflected by greater volumes of international competitors emerging in the offshore market such as China, the Philippines, Malaysia, Vietnam, Hungary and Poland which signals that firms are considering new alternative markets when looking to offshore, rather than established offshore destinations such as India (e.g. Marriott, 2006).

This is because there is a belief that the nature of the IT industry has vastly matured over the past two decades, which has allowed organizations to view IT as a commodity rather than a service (e.g. Carr 2003) resulting in firms being able to choose their offshore vendor regardless of location.

Also, as the primary driver for firms to offshore is cost reduction, the increasing labour costs in places such as India

(Table 2) has resulted in firms looking for cheaper, alternative locations to offshore their IT services to.

The emerging of new offshore locations in places such as Asia, can also be explained by observing the increase in wages in traditionally strong offshore markets such as India which has resulted in firms looking for alternative, more cheaper locations to offshore such as Asia.

Also, it is suggested that the vendor in a specific country you offshore to does not guarantee that proper planning, management and execution will be provided, and the high levels of failure experienced in established offshoring locations prove that (McCue 2005) reputation of a particular country does not guarantee a higher level of success for firms who offshore.

Aron and Singh (2005) further back this argument up by stating that companies who are looking to offshore their services are making three critical mistakes.

Firstly, they state that firms are focusing too much of their effort on choosing the right location and negotiating prices, but not enough time is spent on which processes they should offshore and what they shouldn't.

This implies that businesses put too much emphasis on the actual offshore destination and the cost benefits they may receive from a particular location, rather than taking the time to properly plan, schedule and execute the offshoring agreement.

Secondly, they state that organizations don't take into account all the risks associated with offshoring, such as the transfer of power to the vendor after the negotiation has been completed which may result in unrealistic demands set forth by offshore vendors.

This second point is also backed up by Stringfellow et al (2008) who state that offshoring decisions are made based on visible costs such as labour costs, and do not take into consideration invisible costs such as morale of staff & customer loyalty which may result in lower client retention rates, and higher staff turnover.

Finally, Aron and Singh (2005) point out that firms do not realize that there are a variety of different options and choices available to firms, rather than it being an all-or nothing approach to offshoring.

What this means is that firms must be able to distinguish what should and should not be offshored, and create a balanced approach where certain functions may stay in-house, whereas other functions may be deemed beneficial if outsourced offshore.

In summary, their main argument is that companies who are looking to offshore must get their strategies right to ensure that the offshoring of services results in success, rather than focusing too much on location, and cost benefits

In addition to the arguments set forth by Aron and Singh, Kliem (2004) also states that the benefits of offshoring will not be achievable by firms if risks associated with offshoring are not taken into consideration.

Kliem (2004) suggests that risks associated with financial, technical, managerial, behavioral, legal risks must be considered by firms when offshoring IT projects to reduce the chance of failure.

These views combined paint a picture which shows us that the success of offshoring is heavily dependent on the preparation, planning and experience of the firm wishing to offshore, and the type of benefits the offshore vendor can provide based on its own preparation, planning and experience.

Also, the perception of IT is dramatically changing as stated by scholars such as Carr, who consider IT not as a service, but as a commodity and believe that IT services and development will soon be delivered in similar conditions which are observed in factory lines as seen in the manufacturing industry.

These views suggest that reduction in costs, which is the primary goal of firms wishing to offshore, will only be achievable if the risks associated with offshoring are flushed out, and addressed properly as stated by Kliem.

## IV. CRITICAL ANALYSIS & OPINION

In Sections II & III, we highlighted the main arguments for and against whether or not the location of a particular offshore vendor plays an important role in the overall offshoring process.

Some researchers state that the location of the offshore vendor does matter, as the country you offshore to must fulfill the three major driving factors for offshoring as discussed earlier in this paper.

However, others disagree by stating that focusing on the location of the vendor is a critical mistake by firms, and they should focus more on the planning, management and execution of the offshoring process.

Although I agree that location is somewhat important in the offshoring process, I believe it should not be considered as the major factor by firms when deciding an offshore vendor.

The reasons why I believe this is the case are because of the following three reasons which I have drawn based on critically analyzing both sides of the argument:

Firstly, although I agree that certain countries do provide better cost reductions due to cheaper labour costs, firms must also realize that without the right outlay and planning, firms may face additional costs associated with offshoring (Overby 2003).

Invisible costs that stem from cultural differences, issues in regards communication & interaction as well as ineffective management (Stringfellow et al 2008) affect all offshore vendors regardless of location, and there has even been suggestions that locational and/or vendor diversification (Erber & Sayed-Ahmed 2005) is required to control and manage the risks which stem from offshoring.

Secondly, although continuous work flow & customer service can only be achieved by choosing a specific country that works in a particular time zone or area, this is only possible if the right communication & managerial structure is in place.

Productivity lag which result from cultural differences faced between the offshore vendor and buyer often occurs during IT offshoring, and there have been research to suggest that these differences in culture result in an average 20 percent decline in application development efficiency during the first two years of a contract for organizations going offshore. On top of this decline due to cultural differences, higher than anticipated staff turnover faced by offshore service providers further diminish productivity (Erber & Sayed-Ahmed 2004).

This implies that globally dispersed projects create additional challenges compared to local ones (Carmel 1999), and these challenges are not unique for just one geographical location, but all locations.

Thirdly, although certain countries possess highly skilled IT professionals, IT organizations still face significant declines in quality and service levels, which has led to overall backlash by both internal stakeholders as well as external clients (Joshi & Mudigonda 2008).

Although I disagree with the viewpoint that IT is becoming more of a commodity rather than a skill or service as stated by Carr etc., we cannot deny that the IT industry in general has matured into a phase where IT skills are readily available, and the numbers of companies that are able to provide IT as a service are on the increase.

Furthermore, there is evidence to suggest that organizations will soon have an even larger set of offshore options to choose from as the global IT labor supply grows and matures, which will even further reduce the importance of the geographical location an offshore vendor is from (Carmel & Agarwal 2002).

Based on the three key factors above, I do believe that offshoring in the future will be less dependent as the industry matures over time, whereas business factors such as proper planning and management will become more important factors when trying to achieve offshoring success.

## V. Conclusion

Through this paper, we examined two contrasting views as to whether location or company matters more when offshoring IT services.

On one hand of the equation, there was evidence set forth which suggested that location did matter based on the grounds that only particular countries were able to fulfill the primary drivers for firms wishing to offshore their IT services.

However, our studies have found that the location of the offshoring provider does not play a critical role in the offshoring process by firm due to the fact that unless companies are able to understand the complexity of offshoring and provide the necessary solutions, it does not matter if the location set forth has a reputation for providing IT services because ultimately different risk factors will overtake any benefits a firm may wish to experience by offshoring.

Although there are arguments to suggest that certain countries are better equipped to facilitate offshoring services, the high level of failure experienced by firms proves that unless proper planning, management and execution is factored in the offshoring process, firms will continue to face various challenges and risks posed by cultural barriers when offshoring.

We understand that there may be limitations to our findings, as we were not able to address several primary factors in this paper such as government policies, regional stability, and infrastructural capacities of a country, quality of education etc.

However, these factors cannot be directly controlled by either the organization or the offshore vendor, and hence we did not include these factors into our research findings.

In this regards, we are able to draw to a conclusion to state that the planning, management and execution associated with offshoring plays a much greater role in ensuring success in offshoring.

In future research, it would also be interesting to see how firms react with the increasing labor costs in more well-known markets such as India, and what type of decisions are made in order to accomplish their primary objective as discussed in this paper – reducing costs.

Also, it would be instructive to examine the growth in emerging offshoring markets such as Asia are able to overcome the barriers currently experienced by firms and what type of approach is taken during the offshoring process.

## APPENDIX: REFERENCES